\pdfoutput=1

\documentclass[11pt,twoside,a4paper,cmspaper,final,collab]{cms-tdr}
\def\svnVersion{166417}\def\cmsCernNo{2012-084}\def\cmsCernDate{\today}\def\cmsMessage{Submitted to Physical Review Letters}

\begin{document}\cmsNoteHeader{EXO-11-096}

\hyphenation{had-ron-i-za-tion}
\hyphenation{cal-or-i-me-ter}
\hyphenation{de-vices}

\RCS$Revision: 114310 $
\RCS$HeadURL: svn+ssh://svn.cern.ch/reps/tdr2/papers/EXO-11-096/trunk/EXO-11-096.tex $
\RCS$Id: EXO-11-096.tex 114310 2012-04-04 18:41:17Z alverson $
\newlength\cmsFigWidth
\ifthenelse{\boolean{cms@external}}{\setlength\cmsFigWidth{0.85\columnwidth}}{\setlength\cmsFigWidth{0.45\textwidth}}
\ifthenelse{\boolean{cms@external}}{\providecommand{\cmsLeft}{top}}{\providecommand{\cmsLeft}{left}}
\ifthenelse{\boolean{cms@external}}{\providecommand{\cmsRight}{bottom}}{\providecommand{\cmsRight}{right}}
\ifthenelse{\boolean{cms@external}}{\providecommand{\cmsAPStablebegin}{\begin{ruledtabular}}}{\providecommand{\cmsAPStablebegin}{}}
\ifthenelse{\boolean{cms@external}}{\providecommand{\cmsAPStableend}{\end{ruledtabular}}}{\providecommand{\cmsAPStableend}{}}
\ifthenelse{\boolean{cms@external}}{\providecommand{\labelitonce}{~ \\ }}{\providecommand{\labelitonce}{}}
\ifthenelse{\boolean{cms@external}}{\providecommand{\labelitagain}{~ \ }}{\providecommand{\labelitagain}{}}
\ifthenelse{\boolean{cms@external}}{\providecommand{\boundaries}{}}{\providecommand{\boundaries}{\hline\hline}}
\ifthenelse{\boolean{cms@external}}{\providecommand{\labelfixDM}{ (a) 90\% CL Limits on DM model parameters. }}{\providecommand{\labelfixDM}{ \subfloat[ 90\% CL Limits on DM model parameters.] }}
\ifthenelse{\boolean{cms@external}}{\providecommand{\labelfixADD}{ (b) 95\% CL Limits on ADD parameters. }}{\providecommand{\labelfixADD}{ \subfloat[ 95\% CL Limits on ADD parameters.] }}
\ifthenelse{\boolean{cms@external}}{\providecommand{\ADDlimits}{ADDlimitsGMET50fb.pdf}}{\providecommand{\ADDlimits}{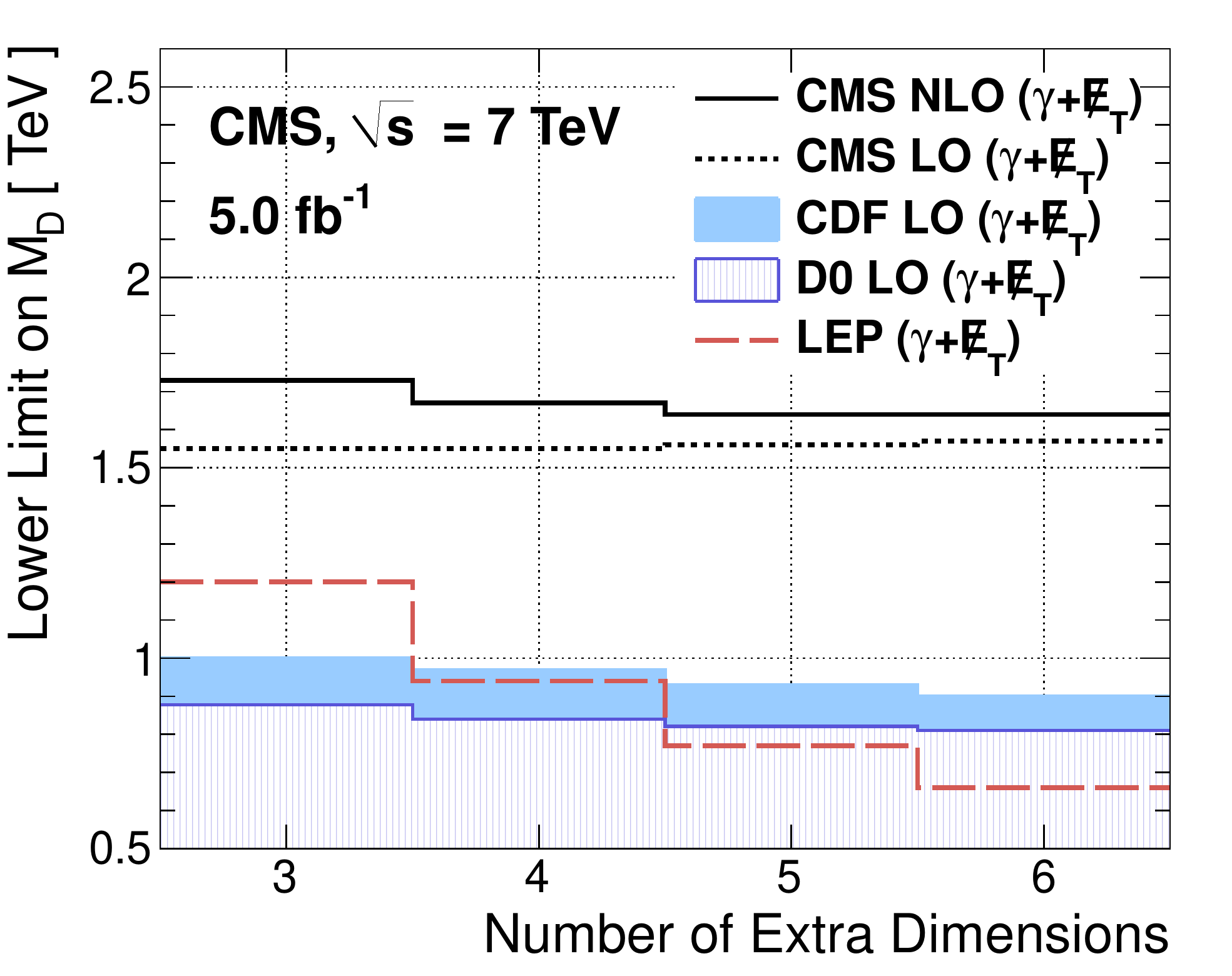}}
\cmsNoteHeader{EXO-11-096} % This is over-written in the CMS environment: useful as preprint no. for export versions
\title{Search for Dark Matter and Large Extra Dimensions in \texorpdfstring{\Pp\Pp}{pp} Collisions Yielding a Photon and Missing Transverse Energy}

\date{\today}

\abstract{
Results are presented from a search for new physics in the final state containing a photon ($\Pgg$) and missing transverse energy (\ETslash ).
The data correspond to an integrated luminosity of  5.0\fbinv collected in \Pp\Pp\ collisions at $\sqrt{s}=7$\TeV by the CMS experiment.
The observed event yield agrees with standard-model expectations for the $\Pgg +\ETslash$ events.
Using models for production of dark-matter particles ($\chi$), we set 90\% confidence level (CL) upper limits of 13.6--15.4\unit{fb} on $\chi$ production in the $\Pgg +\ETslash$ state.
These provide the most sensitive upper limits for spin-dependent $\chi$-nucleon scattering for $\chi$ masses ($M_{\chi}$) between 1 and 100\GeV.
For spin-independent contributions, the present limits are extended to $M_{\chi} <$ 3.5\GeV.
For models with 3--6 large extra dimensions, our data exclude extra-dimensional Planck scales between 1.65 and 1.71\TeV at 95\% CL.
}

\hypersetup{%
pdfauthor={CMS Collaboration},%
pdftitle={Search for Dark Matter and Large Extra Dimensions in pp Collisions Yielding a Photon and Missing Transverse Energy},%
pdfsubject={CMS},%
pdfkeywords={CMS, physics, monophoton, DM, ADD}}

\maketitle %maketitle comes after all the front information has been supplied

Final states in pp collisions at the Large Hadron Collider (LHC), containing a photon ($\Pgg$) of large transverse momentum (\pt) and missing transverse energy (\ETslash ), are used to investigate two proposals of physics beyond the standard model (SM).
One involves a model for dark matter (DM), which is now accepted as the dominant non-baryonic contribution to the matter density of the universe~\cite{DMGeneral}.
Direct searches for a DM candidate ($\chi$) rely on detection through elastic $\chi$-nucleon scattering.
Indirect searches consist of observation of photons or neutrinos produced in $\chi\overline{\chi}$ annihilations in astrophysical sources.
At the LHC, DM can be produced in the reaction $\Pq\Paq \rightarrow  \Pgg \chi\overline{\chi}$, where the photon is radiated by one of the incoming quarks.
The final state is a high-\pt  photon and \ETslash.
Recent theoretical work~\cite{DMTeva, DMLHC} casts this process in terms of a massive mediator in the $s$ channel that couples to a $\chi\overline{\chi}$ pair of Dirac particles.
This process is contracted into an effective theory with a contact interaction scale $\Lambda$, given by $\Lambda^{-2} = g_{\chi}g_qM_M^{-2}$, where $M_M$ is the mediator mass and $g_{\chi}$ and $g_\Pq$ are its couplings to $\chi$ and quarks, respectively.
The model provides a way to connect the $t$-channel $\chi$-nucleon elastic scattering to the $s$-channel pair-production mechanism.
The effective $s$-channel operator can be chosen to represent either a vector or axial-vector, spin-independent or spin-dependent interaction, respectively.

The $\Pgg +$\ETslash\ final state also has sensitivity to models of extra spatial dimensions.
The Arkani-Hamed, Dimopoulos, and Dvali model (ADD)~\cite{ADD}, in particular, provides a possible solution to the hierarchy problem, viz., the disparity between two fundamental scales of nature: the electroweak unification scale ($M_\mathrm{EW}  \approx 100$\GeV) and the Planck scale ($M_\mathrm{Pl} \approx 10^{19}$\GeV).
In this framework, space-time is postulated to have $n$ extra compact spatial dimensions with a characteristic scale $R$, leading to a modified Planck scale, $M_{D}$, given by $M_\mathrm{Pl}^2 \approx M_{D}^{n+2}R^n$.
Assuming $M_{D}$ is of the same order as $M_\mathrm{EW}$, the observed large value of $M_\mathrm{Pl}$ can be interpreted as being a consequence of the ``large" size of $R$ (relative to the Planck length $\approx M_\mathrm{Pl}^{-1}$) and the number of extra dimensions in the theory.
The ADD model predicts the production of gravitons that appear as Kaluza-Klein (KK) modes, where momenta in the extra dimensions appear as observable massive states, except for the zero-mode of the KK excitation, which corresponds to the massless graviton in 4+$n$ dimensions.
The process $\Pq\Paq \rightarrow \Pgg\cPG$, where the graviton \cPG\ escapes detection, motivates the search for events with single high-\pt isolated photons.
While the individual $\Pq\cPG$ couplings are small, the number of expected KK graviton states is large enough to produce a measurable cross section, making it possible to discover large extra dimensions, or to set lower limits on $M_{D}$ as a function of $n$ and upper limits on the ADD cross section.
The same physical phenomena can be accessed through the single-jet (monojet) production channel~\cite{EXO-11-003, ATLASJET}.

This search uses data collected with the Compact Muon Solenoid (CMS) detector~\cite{JINST}.
The momenta of charged particles are measured using a silicon pixel and strip tracker that is immersed in a 3.8\unit{T} superconducting solenoid, and covers the pseudorapidity range $\abs{\eta}< 2.5$.
The pseudorapidity is $\eta = -\ln[ \tan(\theta/2)]$, where $\theta$ is the polar angle measured relative to the counterclockwise-beam direction.
The tracker is surrounded by a crystal electromagnetic calorimeter (ECAL) and a brass-scintillator hadron calorimeter (HCAL).
Both measure particle energy depositions and consist of a barrel assembly and two endcaps that provide coverage in the range of $\abs{\eta} < 3.0$.
A steel/quartz-fiber Cherenkov forward detector (HF) extends the calorimetric coverage to $\abs{\eta} <$ 5.
Muons are measured in gas detectors embedded in the steel return yoke outside of the solenoid.

The primary background for the $\Pgg+$\ETslash\ signal is the irreducible SM background from $Z\Pgg\rightarrow\Pgn\Pagn\Pgg$ production.
This and other SM backgrounds, including $\PW\Pgg$, $\PW\rightarrow \Pe\Pgn$, $\Pgg$+jet, multijet (referred to as QCD), and diphoton events, as well as backgrounds from beam halo and cosmic-ray muons are taken into account in the analysis.

Events are selected from a data sample corresponding to an integrated luminosity of 5.0\fbinv collected using a two-level trigger system, with Level-1 (L1) seeding High Level Triggers (HLT).
The single-photon triggers comprising this search are not prescaled, and are fully efficient within the selected signal region of $\abs{\eta^{\Pgg}} <1.44$~\cite{CMS-PAS-EGM-10-006} and $\pt^{\Pgg} >145$\GeV.
To optimize the analysis for single high-\pt photons accompanied by large \ETslash, photon candidates are restricted to be in the central barrel region, where purity is highest.
To distinguish photon candidates from jets, we apply additional calorimetric selections.
The ratio of energy deposited in the HCAL to that in the ECAL within a cone of $\Delta R =$ 0.15 is required to be less than 0.05, where $\Delta R = \sqrt{(\Delta\phi)^2 + (\Delta\eta)^2}$ is defined relative to the photon candidate and the azimuthal angle $\phi$ is measured in the plane perpendicular to the beam axis.
Photon candidates must also have a shower distribution in the ECAL consistent with that expected for a photon~\cite{CMS-PAS-EGM-10-006}.

Isolation requirements on photon candidates impose upper limits on the energy deposited in the detector around the axis defined by the EM cluster position and the primary vertex~\cite{CMS-PAS-EGM-10-006}.
In particular, the scalar sum of \pt depositions in the ECAL within a hollow cone of $0.06<\Delta R<0.40$, excluding depositions within $|\Delta\eta | = 0.04$ of the cluster center, must be $<$4.2\GeV + 0.006$\times \pt^{\Pgg}$, the sum of scalar \pt depositions in the HCAL within a hollow cone of $0.15<\Delta R<0.40$ must be $<$2.2\GeV + 0.0025$\times \pt^{\Pgg}$, and the scalar sum of track \pt values in a hollow cone of $0.04<\Delta R<0.40$, excluding depositions that are closer to the cluster center than $|\Delta\eta|=$ 0.015, must be $<$2.0\GeV + 0.001$\times \pt^{\Pgg}$ (with \pt in GeV units).
The vetoes defined by the $|\Delta\eta |$ cutoffs are needed to maintain high efficiency for photons that initiate EM showers within the tracker.
The tracker isolation requirement is based on tracks that originate from the primary vertex.

Since the high luminosity of the LHC yields multiple pp interactions per bunch crossing, there are several reconstructed vertices per event.
The primary vertex is defined as the vertex that corresponds to the largest sum of the squares of the associated track-\pt values.
However, to ensure that photon candidates are isolated from charged particle tracks in events with multiple vertices, the tracker isolation requirement must be passed by all reconstructed vertices, or the event is rejected.

The \ETslash \ is defined by the magnitude of the vector sum of the transverse energies of all of the reconstructed objects in the event, and is computed using a particle-flow algorithm~\cite{CMS-PAS-PFT-10-002}.
The candidate events are required to have \ETslash \ $> 130$\GeV.

All events are required to have the energy deposited in the crystal containing the largest signal within the photon to be within $\pm3$\unit{ns} of the time expected for particles from a collision.
This requirement reduces instrumental background arising from showers induced by bremsstrahlung from muons in the beam halo or in cosmic rays.
Spurious signals embedded within EM showers that otherwise pass selection criteria are eliminated by requiring consistency among the energy deposition times for all crystals within an electromagnetic shower.
Photon candidates are removed if they are likely to be electrons, as inferred from characteristic patterns of hits in the pixel detector, called ``pixel seeds," that are matched to the EM clusters~\cite{EGM-10-004}.
In addition, a veto applied to events that contain muon candidates, including those that do not emanate from the collision point, prevents bremsstrahlung from muons in cosmic rays and the beam halo from being reconstructed as prompt photons balanced by \ETslash.
Finally, events are vetoed if they contain significant hadronic activity, defined by: (i) a track with $\pt>20$\GeV that is $\Delta R > 0.04$ away from the photon candidate, or (ii) a jet that is reconstructed with $\pt > 40$\GeV using the anti-\kt~\cite{Cacciari:2008gp} particle-flow algorithm~\cite{CMS-PAS-PFT-10-002}, within $|\eta|<3.0$ and $\Delta R < 0.5$ of the axis of the photon.

After applying all of the selection criteria, 75 candidate events are found.

Backgrounds that are out of time with the collisions are estimated from data by examining the transverse distribution of energy in the EM cluster and the time-of-arrival of the signal in the crystal with the largest energy deposition.
Templates for anomalous signals~\cite{EGM-10-012}, cosmic-ray muons, and beam halo events are fitted to a candidate sample that has no timing requirement, which reveals that the only significant residual contribution to the in-time sample arises from halo muons, with an estimated $11.1\pm5.6$ events.

Electrons misidentified as photons arise mainly from $\PW\rightarrow \Pe\Pgn$ events.
The matching of electron showers to pixel seeds has an efficiency of $\epsilon=0.9940\pm0.0025$, as estimated with Monte-Carlo simulated events (MC) and verified with $\cPZ\rightarrow \Pe\Pe$ events in data.
Scaling a control sample of electron candidates by $(1-\epsilon)/\epsilon$ yields an estimated contribution of $3.5\pm1.5$ $\PW\rightarrow \Pe\Pgn$ events in the candidate sample.

The contamination from jets misidentified as photons is estimated by using a control sample of EM-enriched QCD events to calculate the ratio of events that pass the signal photon criteria relative to those that pass looser photon criteria but fail an isolation requirement.
Since the EM-enriched sample also includes production of direct single photons, this additional contribution to the ratio is estimated by fitting templates of energy-weighted shower widths from MC-simulated $\Pgg+$jets events to an independent QCD data sample, and used to subtract the $\Pgg+$jets contribution.
This corrected ratio is applied to a subset of the EM-enriched jet events that passes loose photon identification and additional single-photon event selection criteria, providing a background contribution of $11.2\pm2.8$ jet events.

Backgrounds from ($\cPZ\Pgn\Pagn$)$\Pgg$, ($\PW \ell\Pgn$)$\Pgg$, $\Pgg$+jet, and diphoton events are estimated from MC samples processed through the full \GEANTfour-based simulation of the CMS detector \cite{GEANT, GEANTdev}, trigger emulation and event reconstruction used for data.
The $\PW\Pgg\rightarrow \ell\Pgn\Pgg$ samples are generated with \MADGRAPH{}5~\cite{Madgraph}, and the cross section is corrected to include next-to-leading order (NLO) effects through a $K$-factor calculated with \MCFM~\cite{MCFM}.
The $\cPZ\Pgg \rightarrow\Pgn\Pagn\Pgg$, $\Pgg$+jet, and diphoton samples are obtained using the \PYTHIA6.424 generator~\cite{Pythia6} at leading order (LO) and CTEQ6L1~\cite{CTEQ6L1} parton distribution functions (PDF).
The $\cPZ\Pgg \rightarrow\Pgn\Pagn\Pgg$ sample is also scaled up to reflect NLO contributions given in Ref.~\cite{baur2}.
Good agreement between data and the rescaled MC for the $\cPZ\Pgg\rightarrow \ell\ell\Pgg$ channel has been obtained in previous CMS studies~\cite{Vgamma}.
The uncertainty on $\cPZ\Pgg \rightarrow\Pgn\Pagn\Pgg$ and the other backgrounds takes into account several sources: theoretical uncertainties on the LO cross section and $K$-factors; the uncertainty on the scale factor that models the data--MC difference in the efficiency; and systematic uncertainties on the photon-vertex assignment, modeling of pile-up, and the accuracy of the energy calibration and resolution for photons, jets, and \ETslash.
The expected contribution from the $\cPZ\Pgg \rightarrow\Pgn\Pagn\Pgg$ process to the background is $45.3\pm 6.8$ events.
The combined expected background from ($\PW\ell\Pgn$)$\Pgg$, $\Pgg$+jet, and diphoton events is $4.1 \pm 1.0$.

\begin{figure}[ht!]
\begin{center}
\includegraphics[width=\cmsFigWidth]{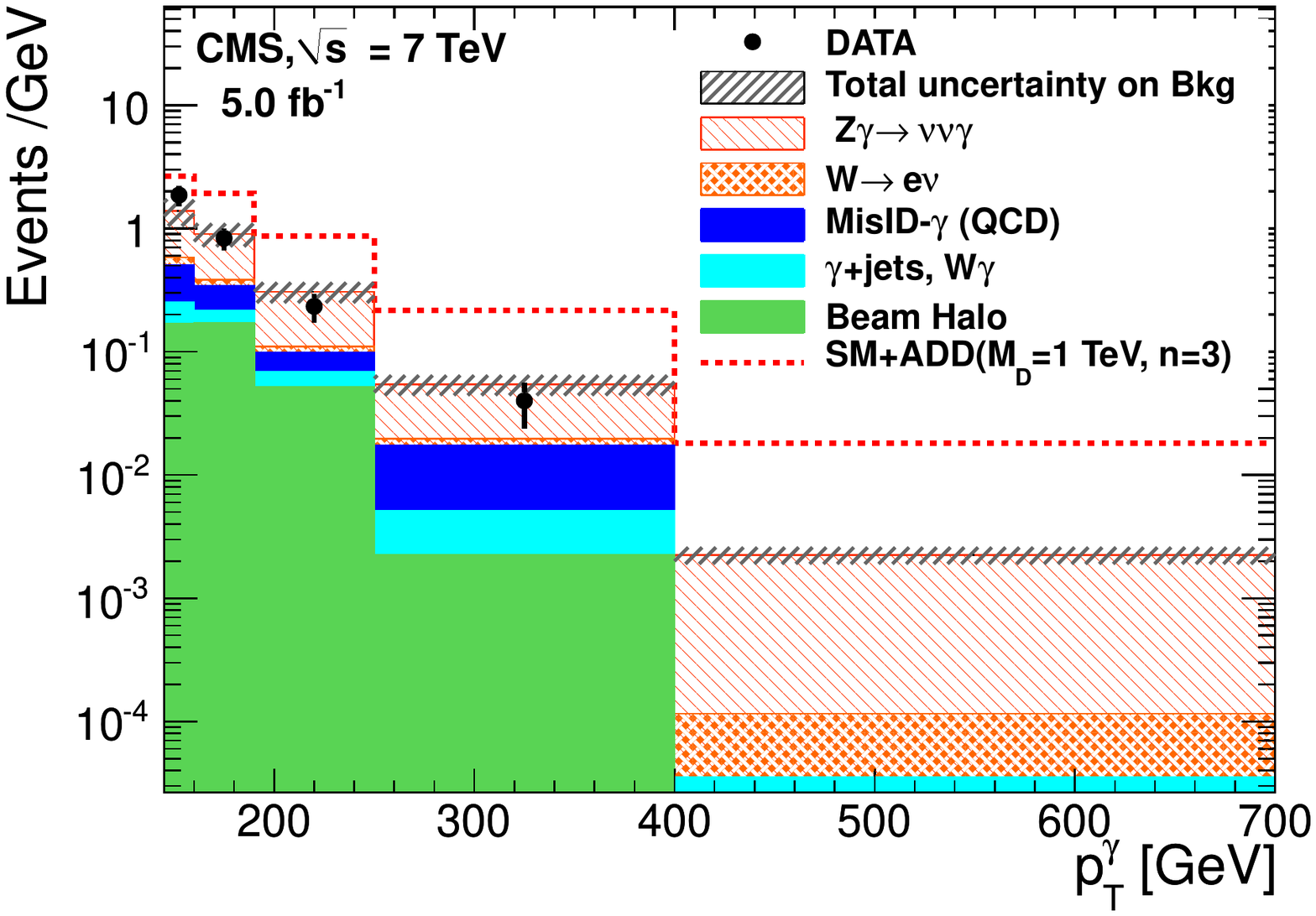}
\caption{The photon \pt distribution for the candidate sample, compared with estimated contributions from SM backgrounds and a prediction from ADD for $M_{D}=1\TeV$ and $n$~=~3.}% ACK Feb20 moved distribution, no more 'of'
\label{fig:stack_plot}
\end{center}
\end{figure}

The 73 observed events in data agree with the total expected background of $75.1\pm9.4$ events.
Distributions in photon \pt for the selected candidate events and for those estimated from background are shown in Fig.~\ref{fig:stack_plot}.
The spectra expected from ADD for $M_{D}$ = 1\TeV and $n$ = 3 are superimposed for comparison.
Based on these results, exclusion limits are set for the DM and ADD models.

The limits on the cross sections are calculated by dividing the difference between the number of events in data and the predicted number of background events by the product $A\times\epsilon\times \mathcal L$, where $A$ is the geometric and kinematic acceptance of the selection criteria, $\epsilon$ is the selection efficiency for signal, and $\mathcal L$ is the integrated luminosity.
$A\times\epsilon$ is calculated by estimating $A\times\epsilon_\mathrm{MC}$ from the MC and multiplying it by a scale factor to account for the difference in efficiency between MC and data.

The efficiency associated with the product $A\times\epsilon_\mathrm{MC}$ for the signal cross section for both models is determined from MC samples.
For the model of DM, the MC samples are produced using a software package from Ref.~\cite{DMLHC}, requiring $\pt^{\Pgg}>$ 125\GeV and $|\eta^{\Pgg}|<$ 1.5.
The estimated value of $A\times\epsilon_\mathrm{MC}$ for $M_{\chi}$ in the range 1--100\GeV is between 30.5--31.0\% for vector and 29.2--31.4\% for axial-vector couplings, respectively.
The spectra for ADD MC events are generated using \PYTHIA8.145~\cite{Pythia8}, requiring $\pt^{\Pgg} >$ 130\GeV, and scaled to NLO using a $K$-factor from Ref.~\cite{kfact}.
The factor $A\times\epsilon_\mathrm{MC}$ for ADD is in the range of 26.5--28.5\% in the parameter space spanned by $n = 3\text{--}6$ and $M_{D} =1\text{--}3\TeV$.

Systematic uncertainties that contribute to the $A\times\epsilon_\mathrm{MC}$ calculation are from the choice of PDF~\cite{PDF4LHC, MSTW08LO, CTEQ6L1}; the selection of the primary vertex for the photon, modeling of pile-up, and the energy calibration and resolution for photons~\cite{CMS-PAS-EGM-10-006}; jets~\cite{JetEnCor2011V2}; and \ETslash~\cite{JME-10-009}.
The total systematic uncertainty on $A\times\epsilon_\mathrm{MC}$ is +4.8\% and $-4.9$\%.

As mentioned above, $A\times\epsilon_\mathrm{MC}$ is multiplied by a scale factor (SF) to account for the difference in efficiency between data and MC.
The calculated SF of $0.90\pm0.11$ combines contributions from the trigger, photon reconstruction, consistency of cluster timing, and vetoes.
The photon HLT is determined to be essentially 100\% efficient for our selection criteria in data and in MC, but is assigned a 2\% uncertainty due to small L1 trigger inefficiencies.
Since the photon identification requirements have similar efficiencies for photons and electrons, the electron efficiency of $0.96\pm0.02$, as measured in $\cPZ\rightarrow \Pe\Pe$ decays is used as the SF.
Corrections for photon reconstruction are described in Ref.~\cite{Vgamma}.
The photon clusters in MC always have consistent timing among individual crystals, and the SF in data is found to be $0.983\pm0.009$ based on a sample of electron events.
The track and jet-veto efficiency is studied in samples of $\PW\rightarrow \Pe\Pgn$ data and MC, and confirmed with $\cPZ\Pgg\rightarrow \Pe\Pe\Pgg$ data.
Since the efficiencies measured in these samples agree within their uncertainties, the SF is set to unity and assigned a systematic uncertainty of $\pm0.10$.
The SF for the cosmic-ray muon veto is determined to be $0.95\pm0.01$ by comparing its efficiency in MC and data in a sample of $\cPZ\rightarrow \Pe\Pe$ events.

\begin{figure}[ht!]
\begin{center}
\includegraphics[width=\cmsFigWidth]{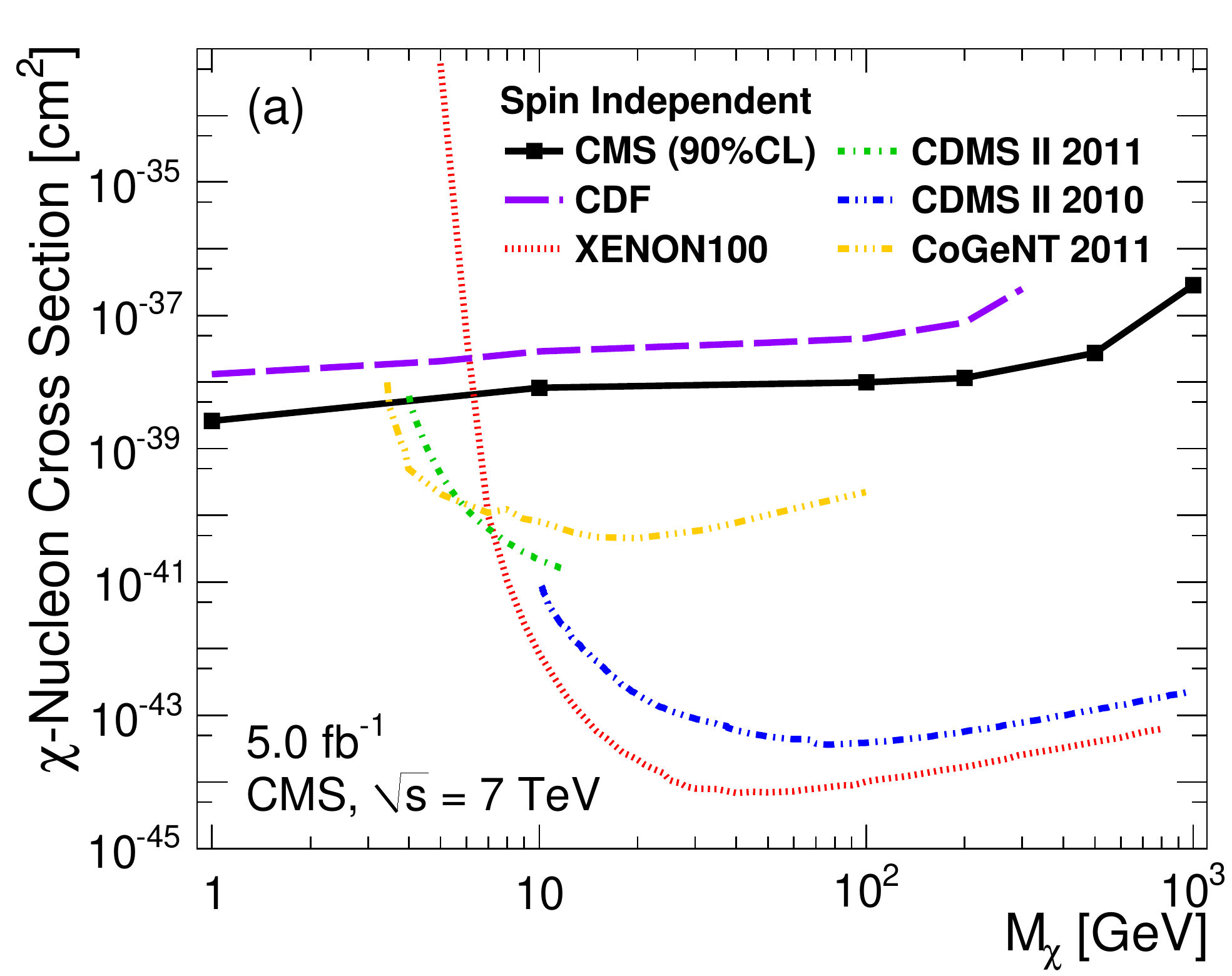}
\includegraphics[width=\cmsFigWidth]{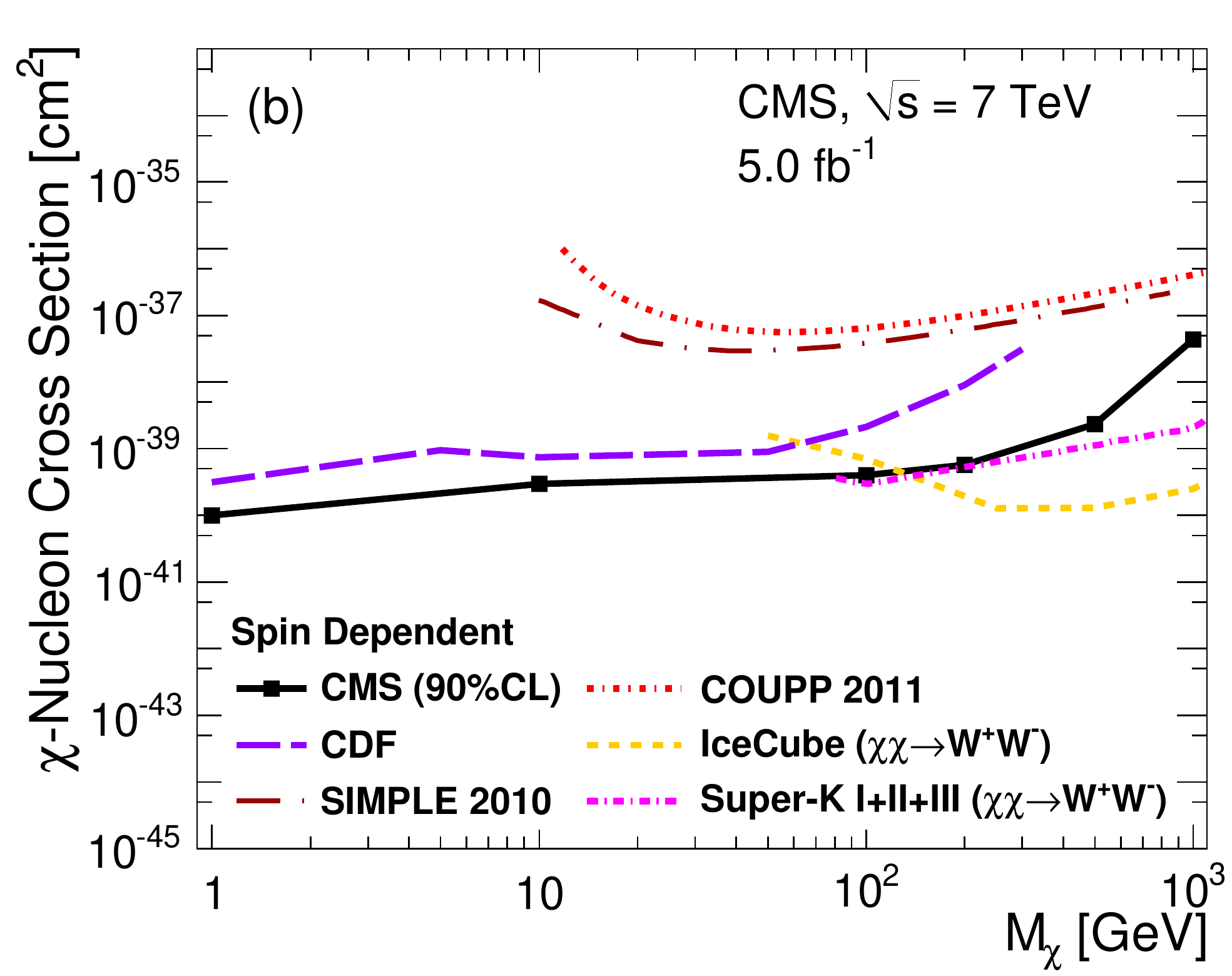}
\caption{The 90\% CL upper limits on the $\chi$-nucleon cross section as a function of $M_{\chi}$ for (a) spin-independent and (b) spin-dependent scattering.  Also shown are the limits from selected experiments with published~\cite{XENON100, CDMS1, CDMS2, COGENT, SIMPLE1, COUPP, IceCube, SUPERK} and preliminary~\cite{CDFdm} results.}\label{fig:DMLimits}
\end{center}
\end{figure}

\begin{table}[ht!]
\centering
\caption{(a) Observed (expected) 90\% CL upper limits on the DM production cross section $\sigma$, and 90\% CL lower limits on the cutoff scale $\Lambda$ for vector and axial-vector operators as a function of the DM mass $M_{\chi}$.  (b) Expected and observed lower limits on $M_{D}$ at 95\% CL, as a function of extra dimensions $n$, with $K$-factors (and without, i.e., $K$ = 1).}
\labelfixDM
{
\cmsAPStablebegin
\begin{tabular}{ccccc}
\boundaries
\multirow{2}{*}{$M_{\chi}$ [\GeVns{}]} & \multicolumn{2}{c}{Vector} & \multicolumn{2}{c}{Axial-Vector} \\
 & $\sigma$~[fb] & $\Lambda$~[\GeVns{}] & $\sigma$~[fb] & $\Lambda$~[\GeVns{}] \\
\hline
1           & 14.3 (14.7) & 572 (568) & 14.9 (15.4) & 565 (561) \\
10 	    & 14.3 (14.7) & 571 (567) & 14.1 (14.5) & 573 (569) \\
100      & 15.4 (15.3) & 558 (558) & 13.9 (14.3) & 554 (550) \\
200      & 14.3 (14.7) & 549 (545) & 14.0 (14.5) & 508 (504) \\
500      & 13.6 (14.0) & 442 (439) & 13.7 (14.1) & 358 (356) \\
1000    & 14.1 (14.5) & 246 (244) & 13.9 (14.3) & 172 (171) \\
\boundaries
\end{tabular}
\cmsAPStableend
\label{tab:LimDM}}
\labelitagain
~ \\
\labelfixADD
{\cmsAPStablebegin
\begin{tabular}{cccc}
\boundaries
\multirow{2}{*}{n}   & \multirow{2}{*}{$K$-factors}  & Expected & Observed \\
 &  & $M_{D}$~[\TeVns{}] & $M_{D}$~[\TeVns{}] \\
\hline
3 & 1.5 & 1.70 (1.53) & 1.73 (1.55) \\
4 & 1.4 & 1.65 (1.53) & 1.67 (1.55) \\
5 & 1.3 & 1.63 (1.54) & 1.64 (1.56) \\
6 & 1.2 & 1.62 (1.55) & 1.64 (1.57) \\
\boundaries
\end{tabular}
\cmsAPStableend
\label{tab:LimADD}}
\end{table}

Upper limits are placed on the DM production cross sections, as a function of $M_{\chi}$, assuming vector and axial-vector operators, summarized in Table~\ref{tab:LimDM}.
These are converted into the corresponding lower limits on the cutoff scale $\Lambda$, also listed in Table~\ref{tab:LimDM}.
The $\Lambda$ values are then translated into upper limits on the $\chi$-nucleon cross sections, calculated within the effective theory framework.
These are displayed in Fig.~\ref{fig:DMLimits} as a function of $M_{\chi}$~\cite{DMTeva}.
The 90\% CL limits are presented in Table~\ref{tab:LimDM}.
Superposed are the results from selected other experiments.
Previously inaccessible $\chi$ masses below ${\approx}3.5\GeV$ are excluded for a $\chi$-nucleon cross section greater than ${\approx}3\unit{fb}$ at 90\% CL.
For spin-dependent scattering, the upper limits surpass all previous constraints for the mass range of 1--100\GeV.
The results presented are valid for mediator masses larger than the limits on $\Lambda$, assuming unity for the couplings $g_{\chi}$ and $g_\Pq$.
The specific case of light mediators is discussed in Ref.~\cite{DMLHC, unitarity}.
The assumptions on $\chi$ interactions made in calculating the limits vary with experiment.
Further, in the case of direct and indirect searches, an astrophysical model must be assumed for the density and velocity distribution of DM.

\begin{figure}[ht!]
\begin{center}
\includegraphics[width=\cmsFigWidth]{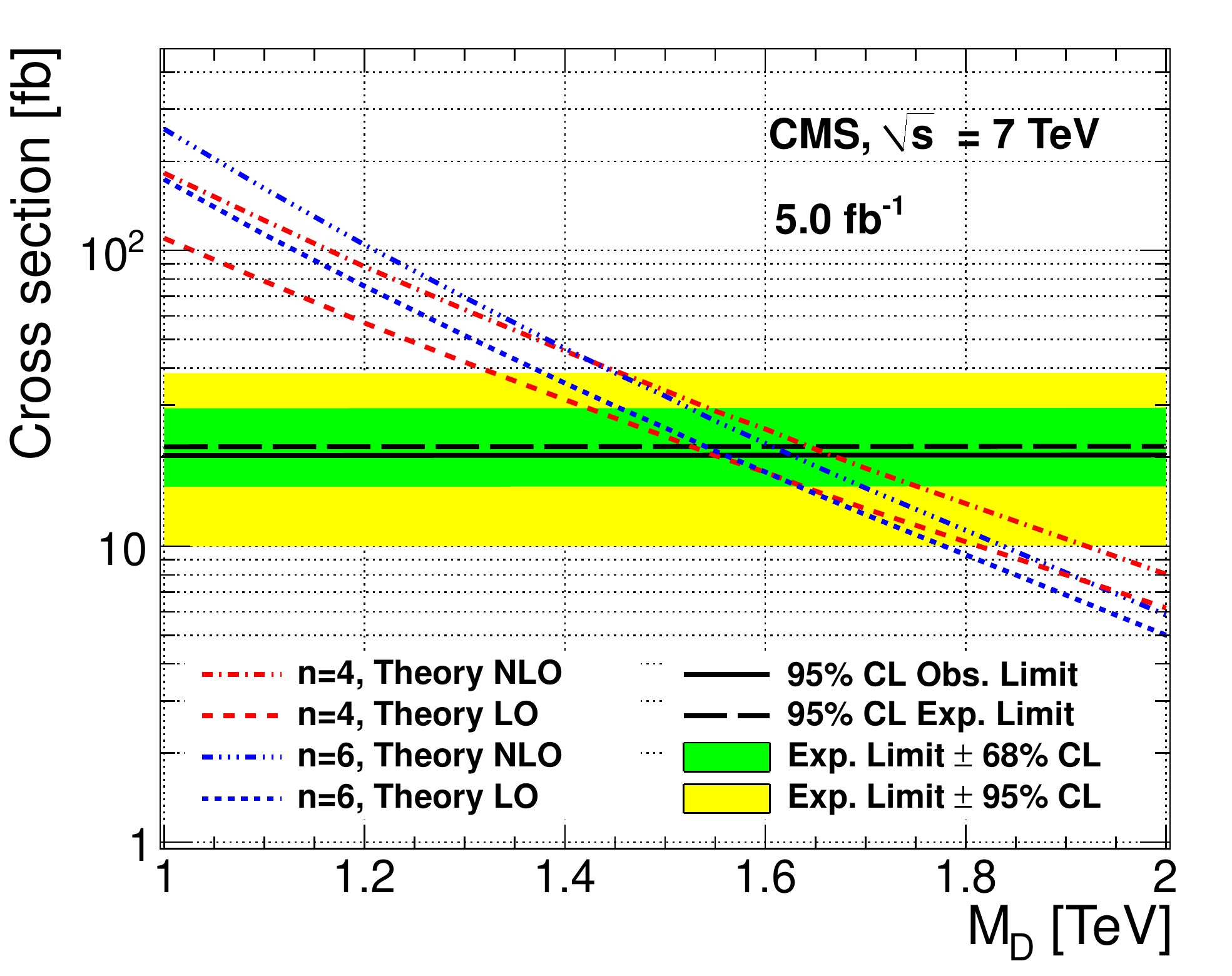}
\includegraphics[width=\cmsFigWidth]{\ADDlimits}
\caption{(a) The 95\% CL upper limits on the LO and NLO ADD cross sections as a function of $M_{D}$ for $n$~=~4 and 6.
(b) Limits on $M_{D}$ as a function of $n$, compared to LO results from similar searches at the Tevatron~\cite{CDFgamma, D0gamma} and LEP~\cite{LEPgamma}.}\label{fig:ADDLimits}
\end{center}
\end{figure}

A set of 95\% confidence level (CL) upper limits are also placed on the ADD cross sections and translated into exclusions on the parameter space of the model.
The upper limits are calculated using a CL$_{\text{s}}$ method~\cite{cls}, with uncertainties parameterized by log-normal distributions in the fit to data.
The limits on $M_{D}$, with and without $K$-factors, are summarized in Table~\ref{tab:LimADD}.
Masses $M_{D} <$ 1.65\TeV are excluded at $95$\% CL for $n=3$, assuming NLO cross sections.
These limits, along with existing LO ADD limits from the Tevatron~\cite{D0gamma, CDFgamma} and LEP~\cite{LEPgamma}, are shown in Fig.~\ref{fig:ADDLimits} as a function of $M_{D}$, for  $n=4$ and $n=6$ extra dimensions.
These results extend significantly the limits on the ADD model in the single-photon channel beyond previous measurements at the Tevatron and LEP experiments, and set limits of $M_{D} > 1.59\text{--}1.66\TeV$ for $n=3\text{--}6$ at 95\% CL.

In summary, the agreement between single-photon production in pp collisions at 7\TeV and standard-model expectations was used to derive significant upper limits on the vector and axial-vector contributions to the $\chi$-nucleon scattering cross section.
This search was complementary to searches for elastic $\chi$-nucleon scattering or $\chi\overline{\chi}$ annihilation.
In addition, through greater sensitivity to the ADD model, the analysis attained the most stringent limits on an effective extra-dimensional Planck scale obtained in the $\Pgg +$\ETslash\ production channel.

\section*{Acknowledgements}
We thank R. Harnik, P. J. Fox, and J. Kopp for help in modeling dark matter production.
We congratulate our colleagues in the CERN accelerator departments for the excellent performance of the LHC machine. We thank the technical and administrative staff at CERN and other CMS institutes, and acknowledge support from: FMSR (Austria); FNRS and FWO (Belgium); CNPq, CAPES, FAPERJ, and FAPESP (Brazil); MES (Bulgaria); CERN; CAS, MoST, and NSFC (China); COLCIENCIAS (Colombia); MSES (Croatia); RPF (Cyprus); MoER, SF0690030s09 and ERDF (Estonia); Academy of Finland, MEC, and HIP (Finland); CEA and CNRS/IN2P3 (France); BMBF, DFG, and HGF (Germany); GSRT (Greece); OTKA and NKTH (Hungary); DAE and DST (India); IPM (Iran); SFI (Ireland); INFN (Italy); NRF and WCU (Korea); LAS (Lithuania); CINVESTAV, CONACYT, SEP, and UASLP-FAI (Mexico); MSI (New Zealand); PAEC (Pakistan); MSHE and NSC (Poland); FCT (Portugal); JINR (Armenia, Belarus, Georgia, Ukraine, Uzbekistan); MON, RosAtom, RAS and RFBR (Russia); MSTD (Serbia); MICINN and CPAN (Spain); Swiss Funding Agencies (Switzerland); NSC (Taipei); TUBITAK and TAEK (Turkey); STFC (United Kingdom); DOE and NSF (USA).

\bibliography{auto_generated}   % will be created by the tdr script.

\cleardoublepage \appendix\section{The CMS Collaboration \label{app:collab}}\begin{sloppypar}\hyphenpenalty=5000\widowpenalty=500\clubpenalty=5000\textbf{Yerevan Physics Institute,  Yerevan,  Armenia}\\*[0pt]
S.~Chatrchyan, V.~Khachatryan, A.M.~Sirunyan, A.~Tumasyan
\vskip\cmsinstskip
\textbf{Institut f\"{u}r Hochenergiephysik der OeAW,  Wien,  Austria}\\*[0pt]
W.~Adam, T.~Bergauer, M.~Dragicevic, J.~Er\"{o}, C.~Fabjan, M.~Friedl, R.~Fr\"{u}hwirth, V.M.~Ghete, J.~Hammer\cmsAuthorMark{1}, N.~H\"{o}rmann, J.~Hrubec, M.~Jeitler, W.~Kiesenhofer, M.~Krammer, D.~Liko, I.~Mikulec, M.~Pernicka$^{\textrm{\dag}}$, B.~Rahbaran, C.~Rohringer, H.~Rohringer, R.~Sch\"{o}fbeck, J.~Strauss, A.~Taurok, F.~Teischinger, P.~Wagner, W.~Waltenberger, G.~Walzel, E.~Widl, C.-E.~Wulz
% ...
\vskip\cmsinstskip
\dag:~Deceased\\
1:~~Also at CERN, European Organization for Nuclear Research, Geneva, Switzerland\\
% ...

\end{sloppypar}
\end{document}